\documentclass[twocolumn,twoside,pre,a4paper]{revtex4}
\usepackage{amsmath}
\usepackage{epsfig}
\usepackage{subfigure}

\begin{document}

\title{Noise-Induced Transition from Translational to Rotational Motion of
  Swarms}

\author{Udo Erdmann} \email{udo.erdmann@physik.hu-berlin.de}

%\homepage[for animations see: ]{http://www.udoerdmann.de}

\author{Werner Ebeling}

\affiliation{Institut f\"ur Physik, Humboldt-Universit\"at zu Berlin,
  Newtonstra{\ss}e 15, 12489 Berlin, Germany}

\author{Alexander S. Mikhailov}

\affiliation{Abteilung Physikalische Chemie, Fritz-Haber-Institut der
  Max-Planck-Gesellschaft, Faradayweg 4-6, 14195 Berlin, Germany}

\date{\today}

\begin{abstract}
  We consider a model of active Brownian agents interacting via a harmonic
  attractive potential in a two-dimensional system in the presence of noise.
  By numerical simulations, we show that this model possesses
  a noise-induced transition characterized by the breakdown of translational
  motion and the onset of swarm rotation as the noise intensity is
  increased. Statistical properties of swarm dynamics in the weak noise limit
  are further analytically investigated.
\end{abstract}

\maketitle

\section{Introduction}

In different natural and social systems, agents form groups characterized by
cooperative coherent motion \cite{BeCoLe00}. Such collective swarm motions
have been observed and investigated in bacterial populations
\cite{HaMa94,BuBe95,BeCoCziViGu97,BrLeBu98,CziMaVi01}, in slime molds
\cite{LeRe91,KeLe93,Na98,RaNiSaLe99}, for the ants \cite{BoThFoDe98,MaDeDe00}
and fish \cite{FlGrLeOl99,PaEd99,PaViGr02}, in the motion of pedestrians
\cite{HeMo95} and for the car traffic \cite{He01}. To describe these
phenomena, various models of collective coherent motion in populations of
self-propelled or self-driven particles have been proposed (see
\cite{OkLe02,MiCa02}). Some of them are formulated in terms of interacting
automata located on a lattice or having continuous coordinates
\cite{ViCzBeCoSh95,Al96,CziVi00,LeRaCo00}. A different class of models is
based on dynamical equations for individual self-propelled particles,
including velocity-dependent negative friction terms to account for their
active motion and assuming that interactions between the particles are
described by some binary potentials. This latter approach has been used to
phenomenologically characterize motion of biological organisms
\cite{Ni94,Ni96a}, individual blood cells \cite{SchiGr93,SchiGr95} and humans
\cite{HeMo95,He97}, and to describe the behavior of physical particles with
energy depots \cite{SchwEbTi98,EbSchwTi99,ErEbSchiSchw99,ErEbAn00}. Effective
dynamical equations with velocity-dependent friction could be approximately
derived for floating particles which propel themselves by releasing surfactant
into the medium \cite{MiMe97}. A continuum approximation to discrete particle
models, hydrodynamical models of active fluids have been proposed
\cite{ToTu95,CsaCzi97,Er99,ToTu98,MiCa02,Er03}.

The common property of all
swarm models is that they show the emergence of coherent collective flows
starting from a disordered state with random velocity directions of individual
particles (agents). Such kinetic transitions have been extensively
investigated for the automata systems \cite{ViCzBeCoSh95,CziBaVi99} and in the
framework of hydrodynamics \cite{ToTu95,ToTu98}. The ordered states of swarms
can represent simple translational motion or be characterized by vortex flows.
Both spatially distributed and localized swarm states are possible. In a
distributed state, the population fills the entire available medium. In
contrast to this, the population forms compact spatial groups in a localized
state. An interesting example of a localized swarm state is a rotating flock
of finite extent, seen in the simulations of a discrete model of
self-propelling particles and described analytically within a continuum active
fluid approximation \cite{LeRaCo00} (see also \cite{MiCa02}). 

Localized states
of swarms may undergo transitions leading to new dynamical regimes, when the
system parameters or the noise intensity are gradually changed. In the
previous publication \cite{MiZa99}, a noise-induced transition from the
localized state with translational motion to a state with incoherent
oscillations without translational motion was investigated numerically and
analytically for a one-dimensional system of interacting self-propelled
particles. In the present article, we extend investigations of this system to
two spatial dimensions. We study here a population of self-propelled particles
interacting via a parabolic interaction potential corresponding to linear
attracting forces between the pairs. In absence of noise, this dynamical
system has two kinds of attractors, corresponding, respectively, to a compact
traveling state of the entire population and to a state where it rotates as a
vortex without any global translational motion. The aim of our study is to
investigate effects of noise on translational swarm motion. We find that the
system is highly sensitive to stochastic forces. When noise is present, the
traveling swarm is a cloud of particles characterized by different dispersions
in the directions parallel and transverse to the direction of translational
motion. Our numerical simulations confirmed by an approximate analytical study
show that the mean-square transverse dispersion of a swarm is proportional to the
square root of the noise intensity, whereas its dispersion along the direction
of motion depends linearly on the noise intensity. Therefore, for weak noises
the swarm looks like a pancake oriented orthogonally to the motion direction.
When the noise is increased, the swarm gradually acquires a more
symmetric shape. For strong noises, we find that the translational
motion of a swarm becomes suddenly impossible and is abruptly replaced by a
rotational regime with a vortex flow. The detailed formulation of the model is
presented in the next section. In Sect.~\ref{sec:numer-simul} we describe the
results of numerical simulations. The statistical properties of a traveling
swarm in the weak noise limit are approximately explained by an analytical
theory which is constructed in Sect.~\ref{sec:weak-noise-limit}. The paper
ends with conclusions and discussion of obtained results.

\section{The model}
\label{sec:model}

We consider a swarm formed by $N$ identical self-propelled particles of unit
mass interacting via an attractive parabolic pair potential. The dynamics of
the system is given by the following set of evolution equations:
\begin{subequations}
  \label{langev-or}
  \begin{eqnarray}
    \label{eq1a}
    \dot{\bf r}_{i}&=&{\bf v}_{i}\\
    \dot{\bf v}_{i}&=&{\bf F}_{i}
    -\frac{a}{N}\sum_{j=1}^{N}\left( {\bf r}_{i}-{\bf r}_{j}\right)
    +{\bf \xi }_{i}(t).
    \label{eq1b}
  \end{eqnarray}
\end{subequations}
The forces ${\bf F}_{i}$ depend on the particle velocity and are introduced to
take into account active motion of particles. We choose them in the form
\begin{equation}
  {\bf F}_{i}=(1-{\bf v}_{i}^{2}){\bf v}_{i}\,,
\end{equation}
so that, in absence of noise and interactions, the particle acquires the unit
velocity $v=1$. Additionally, the particles are subject to stochastic white
forces $\xi _{i}$ of strength $D$ which are independent for different
particles and are characterized by the correlation functions
\[
  \langle {\bf \xi }_{i}(t)\rangle =0\,,\quad
  \langle {\bf \xi }_{i}(t){\bf \xi }_{j}(t^{\prime })\rangle 
  =2D\delta (t-t^{\prime})\delta _{ij}\,.
\]

This model has previously been introduced in \cite{MiZa99}. It is
phenomenological, but rather general because it can be viewed as a normal form
for a population of particles near a supercritical bifurcation corresponding to
spontaneous onset of active motion (see \cite{MiCa02}). In this model,
attractive interactions between particles have an infinite range and grow
linearly with the separation between them. Because we shall consider only
spatially localized states of the population, our results will hold, however,
also for the situations when interactions are characterized by a finite range,
but it is much larger than the mean swarm diameter. It should be noted that,
in a different context, the model~(\ref{langev-or}) has been considered
already by Rayleigh \cite{Ra45}. 

The study of the one-dimensional version of
the model~(\ref{langev-or}) has shown that, as the noise intensity $D$ is
increased, spontaneous breakdown of translational swarm motion takes place
here \cite{MiZa99}. Some statistical properties of translational swarm motion
in the two-dimensional model (1) (with a slightly different choice of
the forces ${\bf F}_{i}$) have subsequently been discussed \cite{SchwEbTi01}.
For the case of two interacting particles ($N=2$), the rotational modes were
described in \cite{ErEbAn00}, where simulations for small rotating clusters
consisting of 20 particles have also been reported. The aim of the present
work is to perform systematic, numerical and analytical, investigations of the
behavior described by this two-dimensional model.

\section{Numerical simulations}
\label{sec:numer-simul}

Numerical integration of equations~(\ref{langev-or}) was performed using the
Euler scheme with the constant time step of 0.001. In all simulations, the
total number of particles was fixed to $N=300$ and the coefficient $a$
specifying the strength of interactions between the particles was set to
$a=100$.  To produce a traveling localized state of the swarm, special initial
conditions were used. At time $t=0$, all particles had identical positions
and velocities and the noise was switched only a little later, at time $t=30$.

Several statistical characteristics of the swarm were monitored during
simulations. The center of mass ${\bf R}$ of the swarm and its mean velocity
${\bf V}${\bf \ } at time $t$ were defined as ${\bf R}(t)=(1/N)\sum_{i}{\bf
  r}_{i}(t)$ and ${\bf V}(t)=(1/N)\sum_{i}{\bf v}_{i}(t)$, respectively.
Because the cloud of traveling particles in the presence of noise was
anisotropic, we also determined its instantaneous mean-square dispersions in
the directions parallel ($S_{\Vert }$) and orthogonal ($S_{\bot }$) to the
direction of its instantaneous mean velocity ${\bf V}$. They were defined as
\begin{subequations}
  \label{eq:deltas}
  \begin{eqnarray}
    S_{\Vert}(t)&=&\frac{1}{NV^{2}(t)}\sum_{i=1}^{N} \left[ \left( {\bf
          r}_{i}(t)-{\bf R}(t)\right) \cdot {\bf V}(t)\right] ^{2}\\
    S_{\bot}(t) &=&\frac{1}{NV^{2}(t)}\sum_{i=1}^{N} \left[ \left( {\bf
          r}_{i}(t)-{\bf R}(t)\right) \times {\bf V}(t)\right] ^{2}.
  \end{eqnarray}
\end{subequations}
Additionally, angular momenta of all particles with respect to the mass center
were determined as
\begin{equation}
  {\bf L}_{i}(t)=\left( {\bf r}_{i}(t)-{\bf R}(t)\right) \times \left( {\bf
      v}_{i}(t)-{\bf V}(t)\right) .\label{eq:L}
\end{equation}

\begin{figure}[tbph]
  \begin{center}
    \subfigure[]{\label{fig:mean_v_t_harmWW}
      \epsfig{file=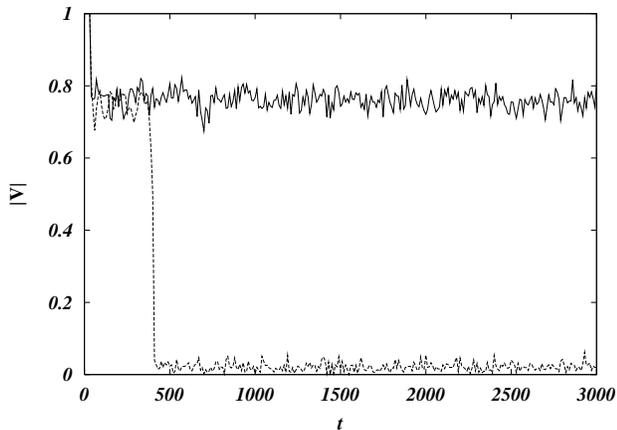,width=8.4cm}}
    \subfigure[]{\label{fig:mean_v_harmWW}
      \epsfig{file=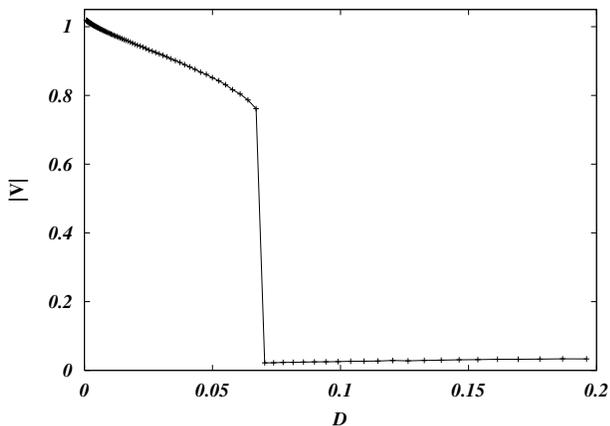,width=8.4cm}}
  \end{center}
  \caption{(a) Time evolution of the mean velocity of a swarm before (solid 
    line, $D=0.067$) and after (dashed line, $D=0.070$) the stochastic
    breakdown of the translational mode. (b) Mean velocity of the swarm with
    increasing noise strength. At a critical noise strength a sharp transition
    in the behavior of the swarm occurs (see also
    Fig.~\ref{fig:snapshot_after}).}
\end{figure}

Figure~\ref{fig:mean_v_t_harmWW} shows the time dependences of the magnitude
$V=\left| {\bf V}\right| $ of the mean swarm velocity for two simulations with
different noises. When the noise is relatively weak ($D=0.067$), its
introduction leads to some fluctuations in the instantaneous swarm velocity
and a decrease of its average level. If a stronger noise ($D=0.07$) is
applied, the swarm velocity first behaves as for the weaker noise, but then
abruptly drops down to a value close to zero. This sudden transition
corresponds to the breakdown of translational motion of the swarm.
In Fig.~\ref{fig:mean_v_harmWW} we have plotted the time-averaged swarm
velocity $\left\langle V\right\rangle $ as the function of the noise intensity
$D$. The average velocity gradually decreases with noise, until the breakdown
occurs at $0.067<D<0.070$.

\begin{figure}[tbph]
  \begin{center}
    \epsfig{file=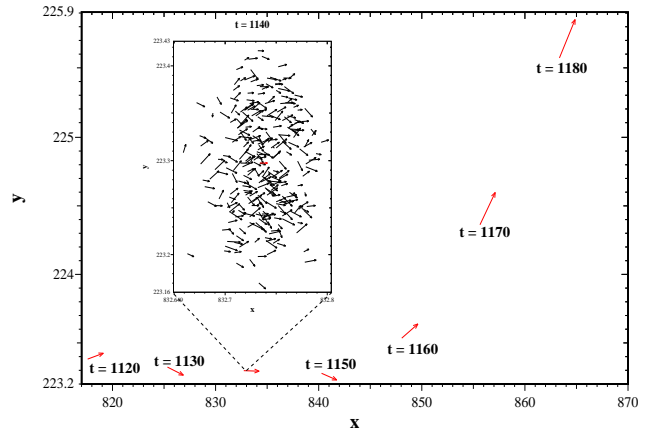,width=8.4cm}
  \end{center} 
  \caption{\label{fig:snapshot_before}
    Motion of the center of mass of the swarm (300 particles) within a certain
    time window for noise below a critical one. For $t=1140$ the corresponding
    snapshot of the swarm is shown. The red arrow shows here th mean swarm
    velocity; the noise intensity is $D=0.067$}.
\end{figure}

In the state with translational motion at relatively weak noises, the
direction of the swarm motion does not remain constant with time. The swarm
travels along a complex trajectory, part of which is shown in
Fig.~\ref{fig:snapshot_before} (such trajectories should correspond to the
Brownian motion of the entire swarm). In the inset in this figure, we display
the distribution of particles in the swarm at some time moment. It can be
noticed that the cloud of particles is significantly squeezed along the
direction of swarm motion.

\begin{figure}[tbph]
  \begin{center}
    \epsfig{file=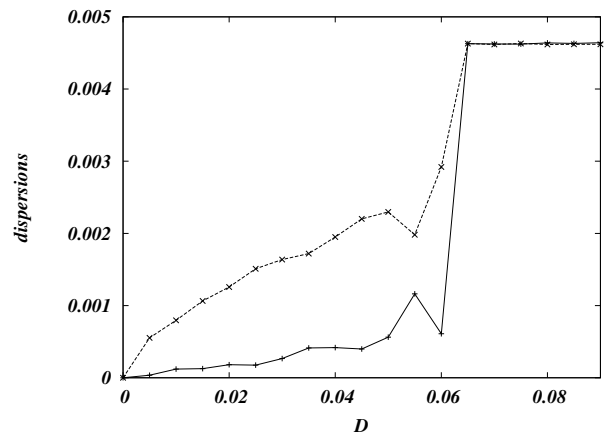,width=8.4cm}
  \end{center}
  \caption{Behavior of the longitudinal (solid line [$+$]) and transversal
    (dashed line [$\times $]) dispersions of the swarm with increasing noise.}
  \label{fig:abweich_harmWW}
\end{figure}

Figure~\ref{fig:abweich_harmWW} shows the computed average longitudinal
($S_{\Vert }$) and transverse ($S_{\bot }$) dispersions as functions of the
noise intensity $D$. For weak noises, $S_{\bot }\gg $ $S_{\Vert }$ so that the
swarm is strongly squeezed. As noise increases, the shape of the swarm becomes
more symmetric and the transversal dispersion approaches the dispersion along
the direction of translational motion. Finally, after the breakdown of
translational motion has taken place for a sufficiently strong noise, the
swarm becomes statistically circular ($S_{\bot }=$ $S_{\Vert }$).

\begin{figure}[tbph]
  \begin{center}
    \epsfig{file=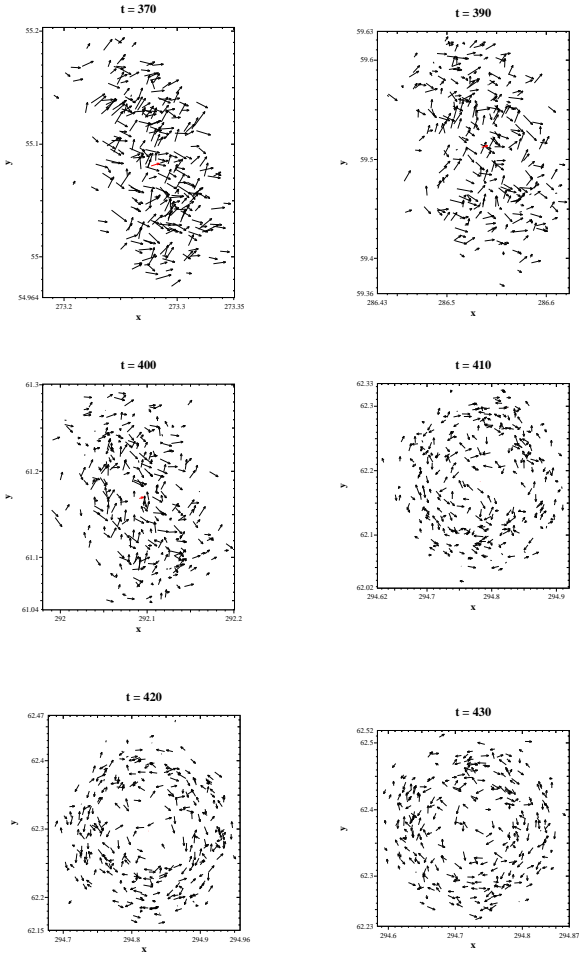,width=8.4cm}
  \end{center}
  \caption{\label{fig:snapshot_after} Several subsequent snapshots of a 
    swarm with 300 particles during the transition from translational motion
    to the rotational mode; the noise intensity is $D=0.070$. }
\end{figure}

The sequence of snapshots in Fig.~\ref{fig:snapshot_after} displays temporal
evolution of the swarm when the noise intensity exceeds the breakdown
threshold. Initially, the swarm is traveling and its shape is similar to that
characteristic for the weaker noises (cf. Fig.~\ref{fig:snapshot_before}).
However, in the course of time the swarm slowly acquires a ring shape, with
particles rotating around its center.  This rotating ring structure
corresponds to a state where translational motion of the entire swarm is
already absent.

\begin{figure}[tbph]
  \begin{center}
    \epsfig{file=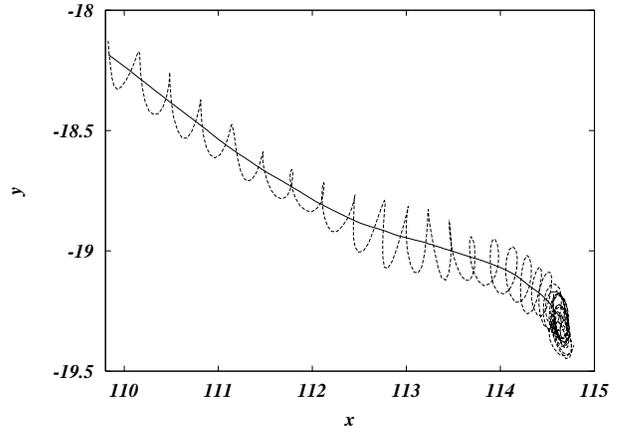,width=8.4cm}
  \end{center}
  \caption{The trajectories of the center of mass (solid line) and of a single
    particle are shown (dashed line) above the critical noise strength (for
    $D=0.070$).}
  \label{fig:trajectories}
\end{figure}

A different visualization of the process, accompanying the breakdown of
translational motion and the transition to a rotating swarm, is chosen in
Fig.~\ref{fig:trajectories}. Here we show the trajectory of motion of the
center of mass of the swarm (solid line) together with the trajectory of
motion of one of its particles (dashed line). We see that, in a traveling
swarm, the particles perform irregular oscillations around its instantaneous
mass center. When the translational motion is terminated and the rotating ring
is formed, such oscillations become transformed into rotations around the ring
center.

\begin{figure}[tbph]
  \begin{center}
    \subfigure[$D=0.067$] {\label{fig:angular_momentum_before}
      \epsfig{file=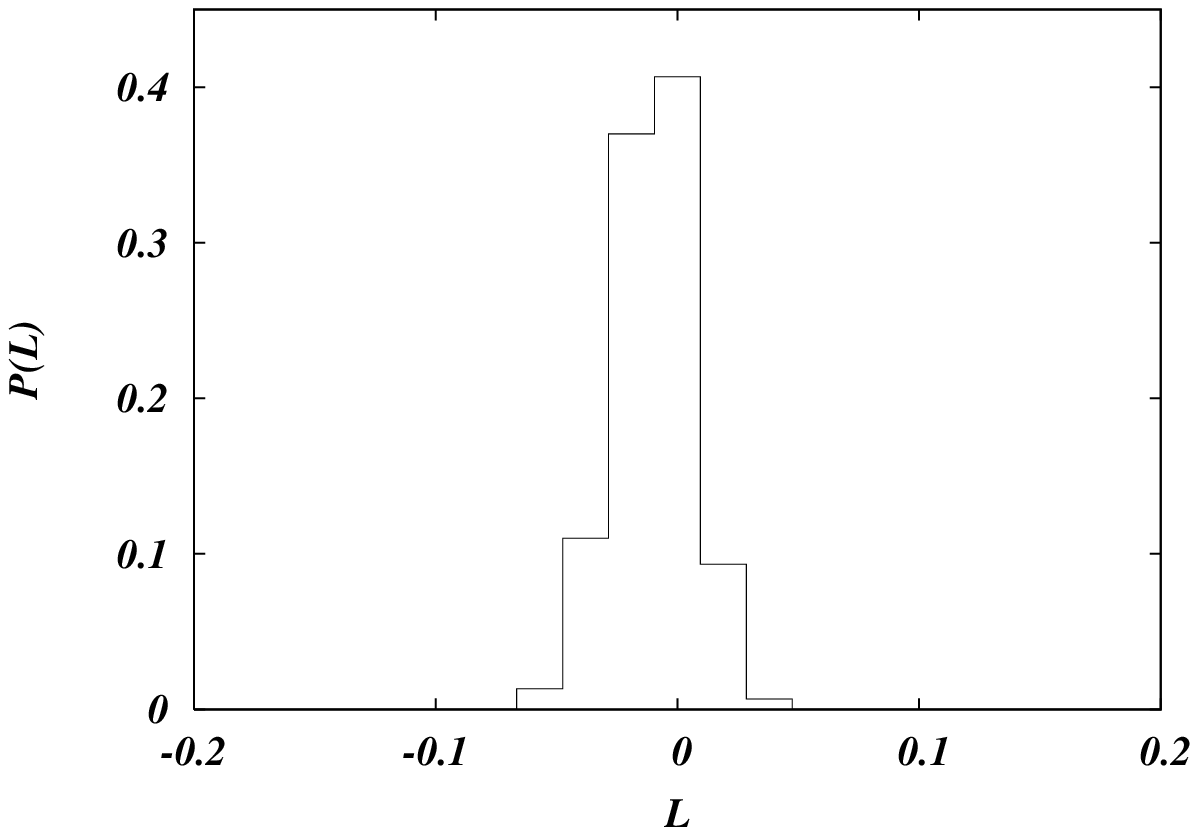,width=4.05cm} }
    \subfigure[$D=0.070$] {\label{fig:angular_momentum_after}
      \epsfig{file=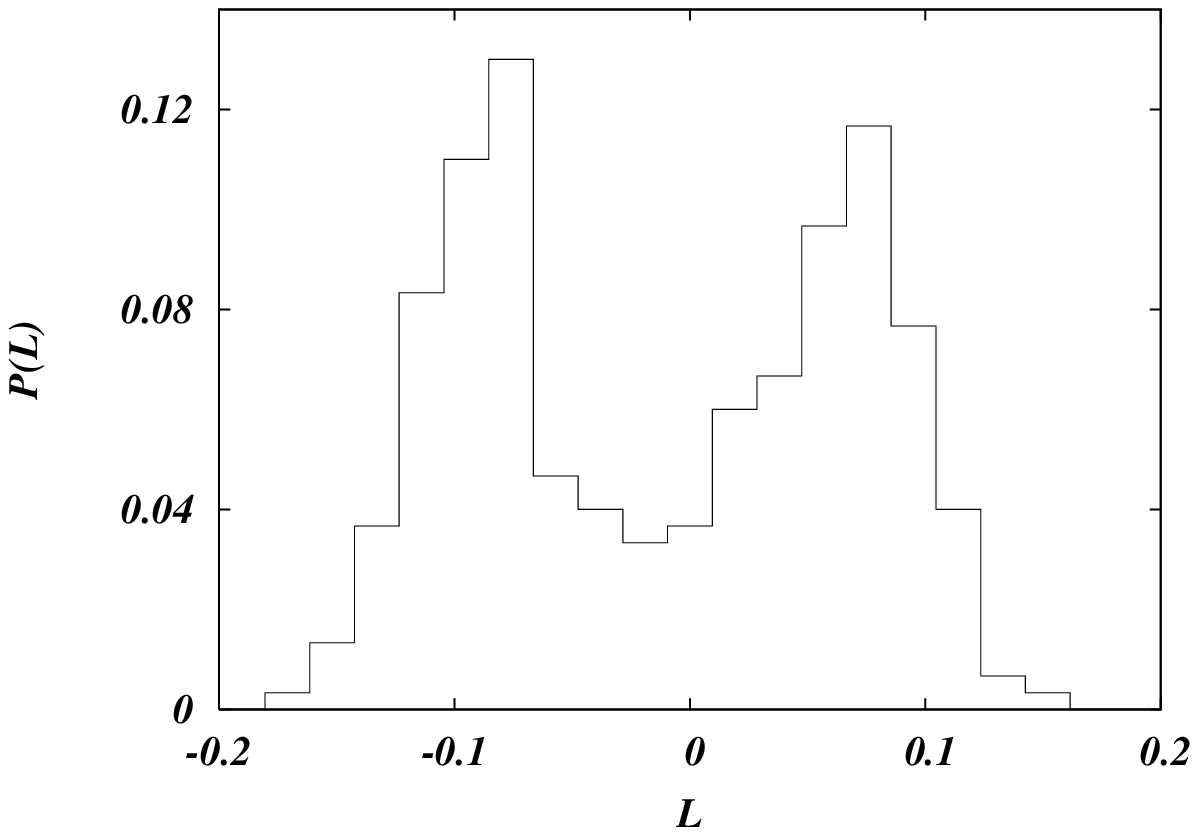,width=4.05cm} }
  \caption{Distribution of angular momenta of the particles (a) in the 
    traveling ($D=0.067$) and (b) rotating ($D=0.070$) swarms.}
  \label{fig:angular_momentum}
  \end{center}
\end{figure}

To provide statistical description of particle motions in the traveling and
rotating states of the swarm, angular momentum distributions $P(L)$ have been
computed. For the state with translational motion, the distribution has a
single central peak at $L=0$ (Fig.~\ref{fig:angular_momentum_before}). In
contrast to this, in the rotational state the distribution has two
symmetrically placed peaks corresponding to a certain non-vanishing momentum
(Fig.~\ref{fig:angular_momentum_after}). Note that the particles inside the
ring are rotating both in the clockwise and counter-clockwise directions, and
the numbers of particles rotating in each direction are approximately equal.
Thus, the swarm does not rotate as a whole and its total angular momentum
remains zero on the average. This behavior is the consequence of the fact that
only long-range attractive interactions between particles are present in the
considered model. It can be expected that, if short range repulsive
interactions are additionally introduced, the breakdown of the rotational
symmetry in the ring would occur and one of the rotation directions would be
selected \cite{ErEbAn00}.

\section{The weak noise limit}
\label{sec:weak-noise-limit}

Our numerical simulations have shown that, for weak noises, the swarm is
strongly squeezed in the direction along its center-of-mass motion and its
longitudinal and transverse dispersions are strongly different. Below we
derive approximate analytical expressions for $S_{\Vert }$ and $S_{\bot }$in
the limit of the small noise intensity $D\rightarrow 0$. 

First, we note that
in this limit the motion of the center of mass of the swarm will remain
approximately linear within very long times or, in other words, the swarm
velocity ${\bf V}$ remains approximately constant on the short time scales
characteristic for the motions of individual particles inside the traveling
swarm. We introduce the coordinate system in such a way that its $x$-axis is
parallel to the direction of the swarm motion and its $y$-axis is orthogonal
to it. The coordinates $x_{i}$ and $y_{i}$ of all particles forming the swarm
can be written as $x_{i}=X+\delta x_{i}$ and $y_{i}=Y+\delta y_{i}$ where $X$
and $Y$ are the coordinates of the swarm center ${\bf R}$. By our choice, we
have $Y=0$, so that $y_{i}=\delta y_{i}$. 

To derive the evolution equation for $X$, we notice that
\begin{equation}
X(t)=\frac{1}{N}\sum_{i=1}^{N}x_{i}(t)\equiv
\left\langle x_{i}(t)\right\rangle\,.
\end{equation}
Summing up the evolution equations for all $x_{i}$, we approximately obtain
\begin{equation}
  \ddot{X}-\left( 1-\dot{X}^{2}\right)
  \dot{X}-3\left\langle \dot{\delta x}_{i}^{2}\right\rangle
  \dot{X}-\left\langle \dot{y}_{i}^{2}\right\rangle \dot{X}=0
  \label{eq:mean_x}
\end{equation}
where we have neglected the terms with the higher powers of velocity
fluctuations $\dot{\delta x}_{i}$ and $\dot{y}_{i}$. 

In the statistical steady
state, $\dot{X}=V={\rm const}$ and (\ref{eq:mean_x}) is reduced to the
equation
\begin{equation}
  \left[ \left( 1-V^{2}\right) -3\left\langle
      \dot{\delta x}_{i}^{2}\right\rangle -\left\langle
      \dot{y}_{i}^{2}\right\rangle \right]V=0\,
\end{equation}
determining the velocity of swarm motion in the presence of noise. Its
solution for the traveling swarm ($V\neq 0$) is
\begin{equation}
  V^{2}=1-3\left\langle
    \dot{\delta x}_{i}^{2}\right\rangle -\left\langle
    \dot{y}_{i}^{2}\right\rangle \,.
  \label{velocity}
\end{equation}

The evolution equation for $\delta x_{i}$ can be obtained by subtracting
equation~(\ref{eq:mean_x}) from the equation for the variable $x_{i}$ in the
model~(\ref{langev-or}). Keeping only the leading terms, linear in deviations
from the mass center, we get
\begin{equation}
  \ddot{\delta x}_{i}+2\dot{\delta x}_{i}+a\delta x_{i}=\xi_{_{i}}^{x}(t)
  \label{eq:lin_fluct_x}
\end{equation}
This is an evolution equation for a damped harmonical oscillator. Note that
fluctuations of $x_{i}(t)$ are not coupled to the transverse component
$y_{i}(t)$.

In a similar way, the evolution equation for the transverse
deviations $y_{i}(t)$ can be obtained,
\begin{equation}
  \ddot{y}_{i}-\left(
    1-V^{2}\right) \dot{y}_{i}+2V\dot{\delta
    x}_{i}\dot{y}_{i}+\dot{y}_{i}^{3}+ay_{i}=\xi _{_{i}}^{y}(t)
  \label{eq:approx_lang_y}
\end{equation}
In this equation, we have retained nonlinear terms. This is done because such
terms are essential for the damping of oscillations of the transverse
component. 

Indeed, if such terms are neglected, we would have
\begin{equation}
  \ddot{y}_{i}-\left( 1-V^{2}\right) \dot{y}_{i}+ay_{i}=\xi _{_{i}}^{y}(t)\,.
\end{equation}
Because, as follows from equation~(\ref{velocity}), we have $V^{2}<1$,
oscillations in $y_{i}$ would then exponentially grow with time. Thus,
nonlinear terms play a principal role for transverse fluctuations and cannot
be neglected even in the weak noise limit, in contrast to the respective terms
for the longitudinal fluctuations $\delta x_{i}$.

As shall be verified at the
end of our derivation, the condition $\langle \dot{y}_{i}^{2}\rangle \gg
\langle \dot{\delta x}_{i}^{2}\rangle $ holds in the weak noise limit.
Therefore, the swarm velocity is mostly influenced by the transverse
fluctuations and we have approximately $V^{2}=1-\left\langle
  \dot{y}_{i}^{2}\right\rangle $. Substituting this into
(\ref{eq:approx_lang_y}), we get
\begin{equation}
  \ddot{y}_{i}-\left(
    \left\langle \dot{y}_{i}^{2}\right\rangle -\dot{y}_{i}^{2}\right)
  \dot{y}_{i}+ay_{i}=\xi_{_{i}}^{y}(t)
  \label{eq:approx_lang_y_final}
\end{equation}

The stochastic differential equation~(\ref{eq:approx_lang_y_final}) does not
include longitudinal fluctuations $\delta x_{i}$ and, furthermore,
fluctuations for different particles $i$ are not coupled here. For subsequent
analysis of this differential equation, we drop the indices and write as
\begin{equation}
  \ddot{y}-\left( \left\langle
      \dot{y}^{2}\right\rangle -\dot{y}^{2}\right) \dot{y}+ay=\xi (t)
  \label{eq:approx_lang_y_final_i}
\end{equation}
with $\langle \xi (t)\xi ^{\prime }(t)\rangle =2D\delta (t-t^{\prime })$. Here
we have assumed that statistical averaging is equivalent to averaging over the
ensemble. 

The approximate solution for the probability distribution of
variable $y$ in the statistical stationary state can be derived for this
equation assuming that the parameter $a$, determining the oscillation
frequency, is large ($a\gg 1$). We introduce slowly varying amplitudes
\begin{equation}
  y(t)= \eta (t)e^{i\omega t}+\eta^{*}(t)e^{-i\omega t} 
  \label{def_eta}
\end{equation}
where $\omega =\sqrt{a}\gg 1$. Substituting this into
(\ref{eq:approx_lang_y_final_i}) and keeping only the resonant terms of the
highest order in $\omega $, we obtain a stochastic evolution equation for the
transverse complex oscillation amplitudes
\begin{equation}
  \dot{\eta}=\omega ^{2}\left\langle \left| \eta
    \right|^{2}\right\rangle \eta -\frac{3}{2}\omega ^{2}\left| \eta \right|
  ^{2}\eta + \zeta (t)
  \label{eq:dot_eta}
\end{equation}
where the complex-valued white noise $\zeta (t)$ has correlation functions
\[
\langle \zeta (t)\rangle =0\,,\quad \langle \zeta (t)\zeta (t^{\prime
})\rangle=0\,,\quad \langle \zeta (t)\zeta^{*}(t^{\prime })\rangle
=\frac{D}{2\omega^{2}}\delta (t-t^{\prime }). 
\]

This stochastic Langevin equation corresponds to the following Fokker-Planck
equation for the probability density $P=P(\eta, \eta ^{*},t)$:
\begin{eqnarray}
  \frac{\partial P}{\partial t} &=&-\frac{\partial}{\partial \eta }
  \left[\omega \left( \left\langle |\eta |^{2}
      \right\rangle \eta -\frac{3}{2}|\eta|^{2}\right)\eta P\right] 
  \label{eq:FPE_eta} \\
  &&-\frac{\partial}{\partial \eta ^{*}}\left[ \omega 
    \left( \left\langle|\eta |^{2}
      \right\rangle \eta -\frac{3}{2}|\eta |^{2}\right)
    \eta^{*}P\right] \nonumber \\
  &&+\frac{1}{2\omega^{2}}\,D\,
  \frac{\partial ^{2}P}{\partial \eta \partial\eta^{*}} \nonumber
\end{eqnarray}
The stationary solution $\overline{P}$ of the Fokker-Planck equation reads
\begin{equation}
  \overline{P}=\frac{1}{Z}\exp \left[
    -\frac{\omega ^{4}}{D}\left(-4\left\langle \left| \eta
        \right| ^{2}\right\rangle \left| \eta
      \right|^{2}+3\left| \eta \right| ^{4}\right) \right]
  \label{prob}
\end{equation}
where the normalization constant $Z$ is given by
\begin{equation}
  Z = \int \exp \left[ -\frac{\omega ^{4}}{D}\left( -4\left\langle \left|
          \eta\right| ^{2}\right\rangle \left| \eta \right| ^{2}+3\left| \eta
      \right|^{4}\right) \right] d^{2}\eta . 
\end{equation}
When the probability distribution is known, the second statistical moment can
be calculated as
\begin{equation}
  \left\langle |\eta |^{2}\right\rangle =\int
  |\eta |^{2}\overline{P}(\eta,\eta ^{*})d^{2}\eta
  \label{eq:dispersion_eta}
\end{equation}
Because the stationary probability distribution (\ref{prob}) depends on
$\left\langle |\eta |^{2}\right\rangle ,$ this is an equation which should be
solved to determine this statistical moment. Let us substitute $\eta =\rho
e^{i\phi }$ and $\omega ^{2}=a$. Then $|\eta|=\rho $ and $d^{2}\eta =\rho
d\rho d\phi $. Therefore, equation~(\ref{eq:dispersion_eta}) takes the form
\begin{equation}
  \left\langle \rho ^{2}\right\rangle
  =\frac{\int_{0}^{\infty }\rho ^{3}\exp {\left[
        -\frac{a^{2}}{D}\left( -4\left\langle \rho ^{2}\right\rangle
          \rho^{2}+3\rho ^{4}\right) \right] }d\rho}
  {\int_{0}^{\infty }\rho \exp {\left[ -\frac{a^{2}}{D}\left(
          -4\left\langle \rho ^{2}\right\rangle \rho
          ^{2}+3\rho^{4}\right) \right] }d\rho }
\label{eq:FPE_eta_subs}
\end{equation}
Introducing the variable $u=\rho \left\langle |\eta
  |^{2}\right\rangle^{-1/2},$ equation (\ref{eq:FPE_eta_subs}) is transformed
to
\begin{equation}
  1=\frac{\int_{0}^{\infty }u^{3}\exp {\left[ -\nu
        \left(-4u^{2}+3u^{4}\right) \right] }du}{\int_{0}^{\infty
    }u\exp {\left[ -\nu\left( -4u^{2}+3u^{4}\right) \right] }du}
  \label{equation_u}
\end{equation}
where 
\begin{equation}
  \nu =\frac{a^{2}}{D}\left\langle |\eta |^{2}\right\rangle
  ^{2}\,.\label{equation_nu}
\end{equation}
Numerical solution of equation (\ref{equation_u}) yields $\nu \simeq 0.22$.
When $\nu $ is known, equation (\ref{equation_nu}) determines
$\left\langle|\eta |^{2}\right\rangle $ as
\begin{equation}
  \left\langle |\eta |^{2}\right\rangle = \nu^{1/2}D^{1/2}a^{-1}\,.
\label{moment}
\end{equation}
Using the definition (\ref{def_eta}) of variable $\eta $, we find that
$S_{\bot}=\left\langle y^{2}\right\rangle =2\left\langle
  |\eta|^{2}\right\rangle $. Thus, we finally obtain the analytical estimate
for the transverse dispersion of the swarm in the weak noise limit,
\begin{equation}
  S_{\bot}=\kappa \frac{D^{1/2}}{a} \label{powerlaw}
\end{equation}
where the numerical coefficient is $\kappa =2\nu ^{1/2}\simeq 0.94$.

The
longitudinal dispersion $S_{\Vert }=\left\langle \delta x^{2}\right\rangle $ is
approximately determined by the linear stochastic differential equation
(\ref{eq:lin_fluct_x}). A straightforward derivation for $a\gg 1$ yields for
this property the analytical estimate
\begin{equation}
  S_{\Vert }=\frac{D}{2a}.
  \label{long}
\end{equation}
We see that in the limit $D\rightarrow 0$ the longitudinal dispersion is
indeed much smaller than the transverse dispersion of the traveling swarm, as
assumed in the above derivation. 

\begin{figure}[tbph]
  \begin{center}
    \subfigure[]{\label{fig:transv_dispersion_scaling}
      \epsfig{file=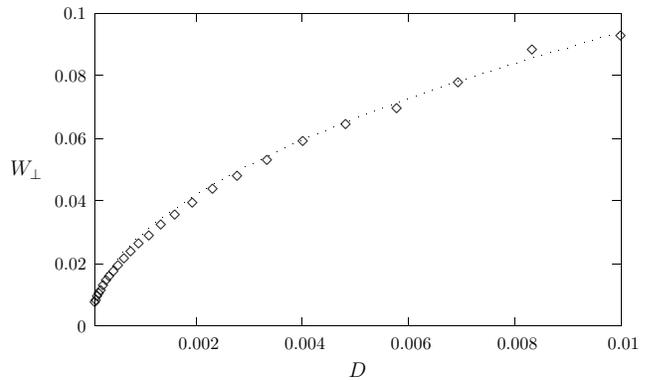,width=8.4cm}
    }
    \subfigure[]{\label{fig:longit_dispersion_scaling}
      \epsfig{file=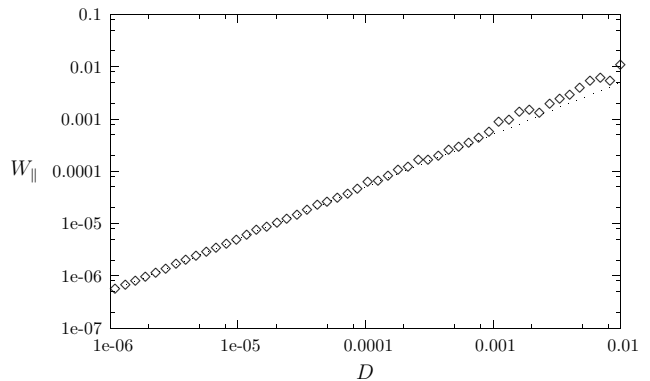,width=8.4cm}
    }
  \end{center}
  \caption{Dispersions of swarm velocities in (a) transversal and 
    (b) longitudinal directions as functions of the noise intensity. The
    symbols show the simulation data. The dotted lines are the theoretically
    predicted power law dependences. }
  \label{fig:dispersion_scaling}
\end{figure}

Note that, for $a\gg 1$, statistical dispersions of transverse and
longitudinal velocity fluctuations $W_{\bot }=\left\langle
  \dot{y}^{2}\right\rangle $ and $W_{\Vert}=\left\langle \delta
  \stackrel{.}{x}^{2}\right\rangle $ are $W_{\bot }=aS_{\bot }$ and $W_{\Vert
}=aS_{\Vert }$. Therefore, in the considered weak noise limit they are given
by $W_{\bot}=\kappa D^{1/2}$ and $W_{\Vert }=D/2$. Comparing these analytical
estimates with the simulations in Fig.~\ref{fig:dispersion_scaling}, we find
that they agree well with the respective numerical data. For the transverse
dispersion, the agreement is found in a wide interval of noise intensities.
The analytical expression for the longitudinal dispersion holds, on the other
hand, only for very weak noises.

\section{Conclusions}\label{sec:conclusions}

We have studied statistical properties of localized swarms with long-range
attractive interactions in two-dimensional media. Our numerical simulations
show that the swarm is highly sensitive to the action of noise. Even very weak
noises lead to strong dispersion of the swarm along the direction orthogonal to
the direction of its translational motion.  The approximate analytical theory
predicts that the transverse dispersion of aswarm increases as $\sqrt{D}$ with
the noise intensity $D$, whereas its longitudinal dispersion depends linearly
on $D$ in the limit $D\rightarrow 0$ and remains therefore much smaller in this
limit. Hence, for weak noises the traveling swarm is strongly squeezed along
the direction of its mass motion. This analytical result is confirmed by
numerical simulations. 

Increasing the noise intensity $D$, we find that translational motion breaks
down when a certain critical intensity is reached.  After the breakdown, the
translational motion is stopped and instead the swarm goes into a rotational
mode where the center of mass of the swarm shows only weak random motion. This
behavior resembles the breakdown of translational motion which was previously
seen for the one-dimensional system \cite{MiZa99}. In contrast to the
one-dimensional case, we could not however analytically treat this transition,
because of the strong fluctuations in the transverse direction.

Though our results are obtained in the model with harmonic attractive
interactions, they are also be applicable for the models with finite-range
attractive interactions between the particles, provided that the size of a
localized swarm (i.e., the statistical dispersion of the coordinates of its
particles with respect to the mass center) is much smaller than the
interaction radius. In this situation, a harmonic approximation of the
interaction potential can be applied. Moreover, similar effects can be
expected for swarms in three-dimensional media. It would be interesting to see
whether the discussed behavior is indeed characteristic for real biological
swarms with long-range interactions between individual organisms, such as bird
flocks or fish schools.

\begin{acknowledgments}  
  We are grateful to L.~Schimansky-Geier and D.~Zanette for useful
  discussions. This study has been performed in the framework of the
  Collaborative Research Center ``Complex Nonlinear Processes'' of the
  Deutsche Forschungsgemeinschaft (DFG-SFB 555).
\end{acknowledgments}

\bibliographystyle{apsrev}
\bibliography{bib/allgemein,bib/ameisen,bib/brown,bib/bakterien,bib/gbt,bib/cells,bib/coherent,bib/erdmann,bib/schleimpilze,bib/toda}
\end{document}